# UiO : **Department of Informatics**
University of Oslo

# Cybersecurity Playbook Sharing with STIX 2.1

A Nested Property Extension for the Course of Action SDO v3.0

Authors: Vasileios Mavroeidis and Mateusz Zych
August 2022

Joint Research Publication by:
University of Oslo and Fovea Research

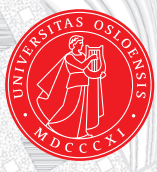

# Table of Contents

# Glossary

| | |
|---|---|
| BPMN | Business Process Model and Notation |
| CACAO | Collaborative Automated Course of Action Operations |
| COA | Course of Action |
| CTI | Cyber Threat Intelligence |
| IACD | Integrated Adaptive Cyber Defense |
| JSON | JavaScript Object Notation |
| OASIS Open | Organization for the Advancement of Structured Information Standards |
| U.S. DOD | United States Department of Defense |
| SCO | STIX Cyber-observable Object |
| SDO | STIX Domain Object |
| SRO | STIX Relationship Object |
| STIX | Structured Threat Information Expression |
| TAXII | Trusted Automated Exchange of Intelligence Information |
| TC | Technical Committee |

# Acknowledgements

This technical report was partially supported by the research projects CyberHunt (Grant No. 303585 - funded by the Research Council of Norway) and JCOP (Grant No. INEA/CEF/ICT/A2020/2373266 - funded by the European Health and Digital Executive Agency through the Connected Europe Facility program).

The authors would like to thank the OASIS *Collaborative Automated Course of Action Operations* and *Threat Actor Context* Technical Committees that, through the working meeting discussions, influenced the direction of this research.

# 1. Introduction

Understanding that interoperable security playbooks will become a fundamental component of defenders' arsenal to decrease attack detection and response times, it is time to consider their position in structured sharing efforts. This report documents the process of extending Structured Threat Information eXpression (STIX) version 2.1, using the available *extension definition* mechanism, to enable sharing security playbooks, including Collaborative Automated Course of Action Operations (CACAO) playbooks.

# 2. Security Playbooks

Security playbooks document cybersecurity processes and procedures and provide a step-by-step approach to orchestration. For instance, playbooks can guide, coordinate, and speed up specific functions in security operation centers and incident response and ensure compliance with organizational policies and regulatory frameworks. Security playbooks appear in different levels of abstraction and detail, different formats such as machine-readable or prose documents and checklists, are modifiable (can be updated) and are aimed to be reused, and based on their scope, can document processes both that are carried out manually and in an automated manner.

# 3. CACAO Security Playbooks

CACAO is an open playbook standard that defines the schema and taxonomy for creating *Collaborative Automated Course of Action Operations* (CACAO) security playbooks. CACAO playbooks document cybersecurity courses of action in a machine-readable format and are designed to be vendor-agnostic for interoperability. Therefore, with CACAO, we can share security playbooks (knowledge sharing) in a structured and standardized way across organizational boundaries and technological solutions. Figure 1 presents the building blocks of a CACAO playbook. Concisely, a CACAO playbook comprises metadata, workflow steps that integrate logic to control the commands to be performed, a set of commands to perform, targets that receive, process, and execute the commands, data markings that specify the playbook's handling and sharing requirements, and extensions that allow introducing additional functionality. Furthermore, for integrity and authenticity, CACAO playbooks can be digitally signed. CACAO playbooks, among other command types, can encapsulate OpenC2, Sigma, and Kestrel commands to enable interoperability at the command level and require minimal modifications to map to an organization's own environment.

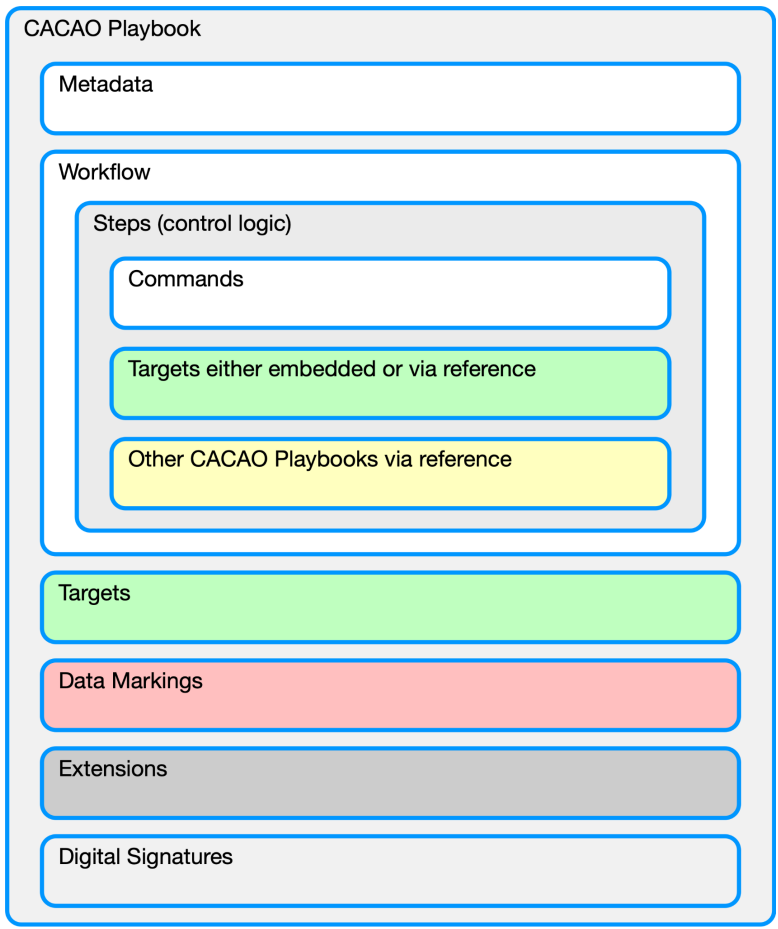

Figure 1. CACAO playbook components [cacao-v1.1].

The CACAO standard is developed and maintained by the OASIS Collaborative Automated Course of Action Operations Technical Committee (CACAO TC)[1].

## 4. Structured Threat Information eXpression

Structured Threat Information eXpression (STIX) is a schema and a language for sharing Cyber Threat Intelligence (CTI). STIX enables organizations to share CTI in a consistent machine-readable manner, allowing security communities to understand better what computer-based attacks they are most likely to see and anticipate and respond to those attacks faster and more effectively [stix-v2.1].

---

[1] https://www.oasis-open.org/committees/cacao

We believe that sharing machine-readable actions to take against threats (i.e., how we respond to a triggered condition) is equally important as sharing CTI that is used to increase our threat situational awareness (i.e., anticipatory threat reduction and detection). Security playbooks complement CTI by providing a set of structured countermeasures that could be performed in the light of an indicator and provide a contextual binding between information about adversaries, their operations, associated technical artifacts, and structured machine-readable response processes.

The STIX 2.1 [stix-v2.1] specification does not provide an option to share machine-readable security playbooks (automated courses of action as per STIX 2.1 terminology), such as CACAO, but allows describing textual courses of action using the Course of Action STIX Domain Object (SDO).

STIX 2.1 and its transport mechanism TAXII 2.1 (Trusted Automated Exchange of Intelligence Information) were approved and released as OASIS standards on June 10, 2021. STIX and TAXII are developed and maintained by the OASIS Cyber Threat Intelligence Technical Committee (CTI TC)[2].

## 5. The Relation Between Security Playbook and Course of Action

There are a plethora of lexical items associated with the concept of *security playbook*. To name a few: cybersecurity playbook, machine-readable security playbook, incident response playbook, course of action, automated course of action, course of action playbook, workflow, orchestration workflow, and runbook.

This report won't go over the lexical semantics of the above terms and their interpretation among security professionals, but for the sake of understanding, this report understands the term security playbook as an equivalent term to course of action playbook.

The U.S. DOD Dictionary of Military and Associated Terms [dod-dictionary] defines *Course of Action as 1) any sequence of activities that an individual or unit may follow, or 2) a scheme*

---

[2] https://www.oasis-open.org/committees/cti

*developed to accomplish a mission.* Such systematic processes, when documented, comprise a playbook.

## 6. STIX 2.1 Course of Action Object

STIX 2.1 defines Course of Action *as an action taken to prevent or respond to an attack. It is aimed to describe technical, automatable responses (applying patches, reconfiguring firewalls) but can also describe higher-level actions like employee training or policy changes* [stix-v2.1_coa].

According to the STIX 2.1 specification [stix-v2.1_coa]:
*Currently, the Course of Action object supports basic use cases, like sharing prose courses of action, and does not support the ability to represent automated courses of action or contain properties to represent metadata about courses of action.*

However, the specification emphasizes that the Course of Action SDO reserves a property (see Table 1) that will be used to capture structured/automated courses of action in a future STIX release.

Table 1. STIX 2.1 - Course of Action SDO - Specific Properties [stix-v2.1_coa].

| Course of Action Specific Properties | | |
|---|---|---|
| `name`, `description`, `action` | | |
| **Property Name** | **Type** | **Description** |
| **type** (required) | `string` | The value of this property **MUST** be `course-of-action`. |
| **name** (required) | `string` | A name used to identify the Course of Action. |
| **description** (optional) | `string` | A description that provides more details and context about the Course of Action, potentially including its purpose and its key characteristics. |
| **action** (reserved) | `RESERVED` | RESERVED – To capture structured/automated courses of action. |

Hereafter, the report discusses the technical aspects of our research. Using the STIX 2.1 *extension definition* mechanism, we extended the Course of Action SDO to support sharing security playbooks and introduced a set of decriptive metadata. The extension is playbook format agnostic and can share any type of playbook, including CACAO.

# 7. Technical Options for Sharing Machine-Readable Security Playbooks Using the STIX 2.1 Course of Action Object

This section describes two technical approaches for sharing machine-readable security playbooks (e.g., CACAO, BPMN) using STIX 2.1 and, in particular, the Course of Action SDO.

**Approach 1:** Utilize the (reserved) property *action* of the Course of Action SDO (see Table 1).

This approach requires updating the STIX specification. As shown in Table 1, a data type for the property *action* should be specified. CACAO playbooks are documented in JavaScript Object Notation (JSON); therefore, the data types *string* or *binary* should be sufficient. However, using one of the existing specification-defined data types restricts the standard to using and sharing only one playbook format, which is a viable solution if, for example, only CACAO playbooks are aimed to be shared.

In the case that more granularity is required, like sharing playbooks in other formats than CACAO (e.g., Ansible, BPMN), producers need to have the option to specify the format their playbook is documented, and consumers need to be able to determine the format of the playbook they received. From a STIX 2.1 implementation perspective, a new data type could be defined (e.g., playbook) and used with the property *action* to represent a special set of key/value pairs where the key defines the playbook format/standard and the value includes the playbook. In addition, an open vocabulary for playbook formats should be introduced.

**Approach 2:** Extend the Course of Action SDO with a set of properties via an *extension definition* [stix-v2.1_extension].

The research documented in this report and the associated technical deliverables are based on Approach 2. In particular, this report introduces a STIX 2.1 extension that augments the Course of Action SDO with properties that allow describing, embedding, and sharing machine-readable security playbooks. The decision to extend the Course of Action SDO and not introduce a new domain object was based on the facts that the terms security playbook and course of action are semantically very close (i.e., security playbook can be a subclass of course of action) and that extending an object allows making use of the existing specification-defined relationships between objects, such as the ones between the Course of Action and other objects (see Table 2). However, according to the STIX 2.1 specification [stix-v2.1], relationships are not restricted to the ones listed and can be created between any objects using the *related-to* or user-defined relationship types.

Table 2. STIX 2.1 - Course of Action SDO - Specific Relationships [stix-v2.1_coa].

| Source | Relationship Type | Target | Description |
|---|---|---|---|
| `course-of-action` | `investigates` | `indicator` | This Relationship describes that the Course of Action can be used to investigate the Indicator. |
| `course-of-action` | `mitigates` | `attack-pattern`, `indicator`, `malware`, `tool`, `vulnerability` | This Relationship describes that the Course of Action can mitigate (e.g. respond to a threat) the related Attack Pattern, Indicator, Malware, Vulnerability, or Tool.<br><br>For example, a `mitigates` Relationship from a Course of Action object to a Malware object indicates that the course of action mitigates the malware. |
| `course-of-action` | `remediates` | `malware`, `vulnerability` | This Relationship describes that the Course of Action can be used to remediate (e.g. clean up) the malware or vulnerability |

## 8. STIX 2.1 Extension Definition

STIX 2.1 includes a mechanism in the form of a STIX object (Extension Definition object) that allows producers of threat intelligence to extend or create new STIX domain, cyber observable, or relationship objects in a standardized manner [stix-v2.1_extension]. The Extension Definition object contains information about the extension and any additional properties and objects it defines. The *schema* property (see Table 3) points to the normative definition of the extension, and it should be a URL pointing to a JSON schema.

Table 3. STIX 2.1 - Extension Definition - Specific Properties [stix-v2.1_extension].

| Extension Definition Specific Properties | | |
|---|---|---|
| `name, description, schema, version, extension_types, extension_properties` | | |
| **Property Name** | **Type** | **Description** |
| **type** (required) | `string` | The `type` property identifies the type of object. The value of this property **MUST** be `extension-definition`. |
| **name** (required) | `string` | A name used for display purposes during execution, development, or debugging. |
| **description** (optional) | `string` | A detailed explanation of what data the extension conveys and how it is intended to be used.<br><br>While the description property is optional this property **SHOULD** be populated.<br><br>Note that the `schema` property is the normative definition of the extension, and this property, if present, is for documentation purposes only. |
| **schema** (required) | `string` | The normative definition of the extension, either as a URL or as plain text explaining the definition.<br><br>A URL **SHOULD** point to a JSON schema or a location that contains information about the schema.<br><br>**NOTE:** It is recommended that an external reference be provided to the comprehensive documentation of the extension-definition. |
| **version** (required) | `string` | The version of this extension. Producers of STIX extensions are encouraged to follow standard semantic versioning procedures where the version number follows the pattern, MAJOR.MINOR.PATCH. This will allow consumers to distinguish between the three different levels of compatibility typically identified by such versioning strings.<br><br>As with all STIX Objects, changing a STIX extension definition could involve STIX versioning. See section 3.6.2 for more information on versioning an object versus creating a new one. |
| **extension_types** (required) | `list of type enum` | This property specifies one or more extension types contained within this extension.<br><br>The values for this property **MUST** come from the `extension-type-enum` enumeration.<br><br>When this property includes `toplevel-property-extension` then the **extension_properties** property **SHOULD** include one or more property names. |
| **extension_properties** `(optional)` | `list of type string` | This property contains the list of new property names that are added to an object by an extension.<br><br>This property **MUST** only be used when the **extension_types** property includes a value of `toplevel-property-extension`. In other words, when new properties are being added at the top-level of an existing object. |

Figure 2 presents, at a high level, the composition of a STIX object type instance that uses an extension. The blue rectangle represents an instance of a new or extended STIX domain, cyber-observable (SCO), or relationship (SRO) object type. The red rectangle represents the associated Extension Definition object (see Table 3) that provides information about the new or extended object type. The amber rectangle captures the normative definition of the extension and it is a JSON schema.

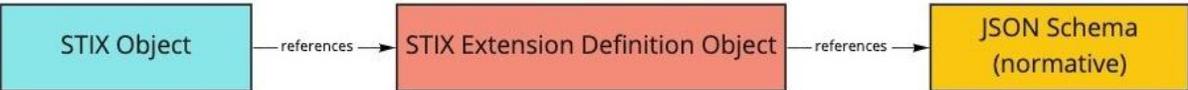

Figure 2. High-level composition of a STIX object type instance that uses an extension.

The *extension definition* can extend STIX in three different ways.

1. Define new STIX object types such as SDO, SCO, or SRO. Figure 3 presents an instance of a new SDO type and its *extension definition* via the Extension Definition object.

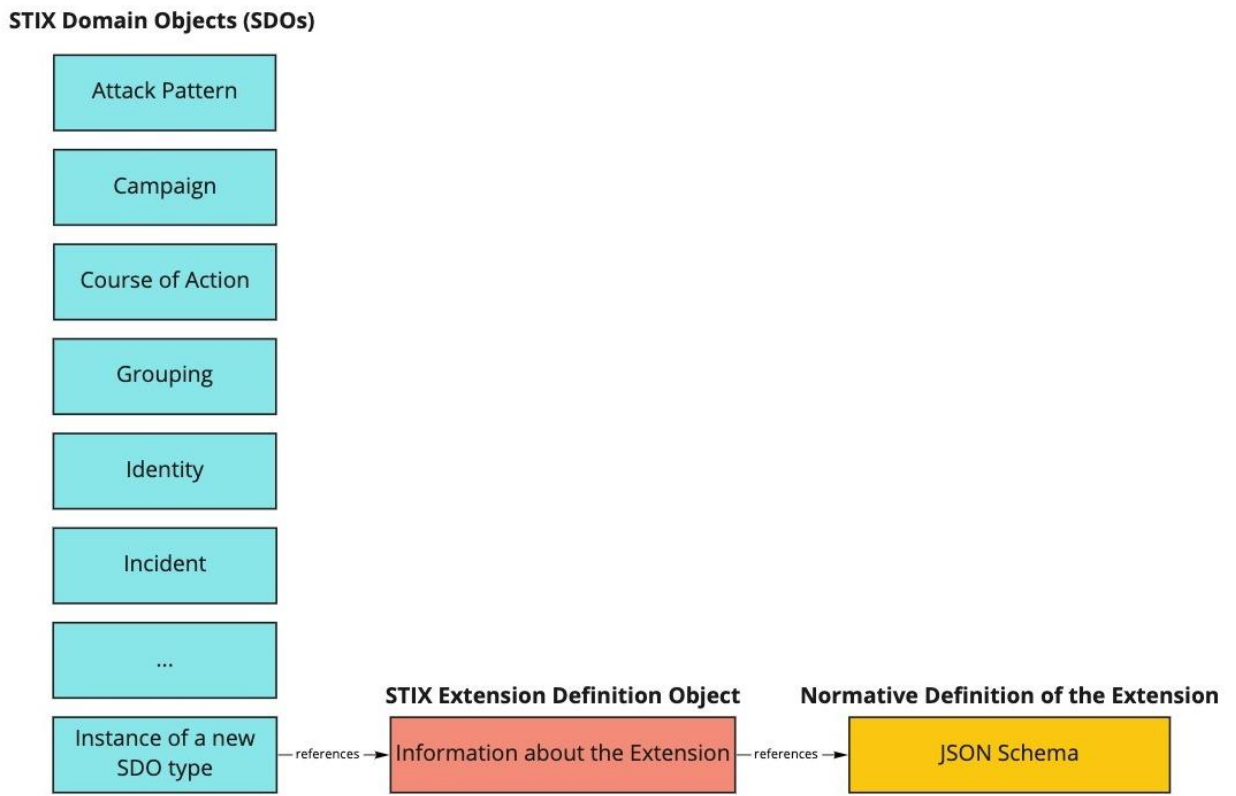

Figure 3. Instance of a new SDO type and its associated extension definition.

2. Define additional properties for an existing STIX Object type as a nested property extension. This is typically done to represent a sub-component or module of one or more STIX Object types. This research utilizes the same approach (nested property extension) to extend the Course of Action SDO type to support describing, embedding, and sharing security playbooks, including structured approaches like CACAO. Figure 4 presents a Course of Action instance with a nested property extension.

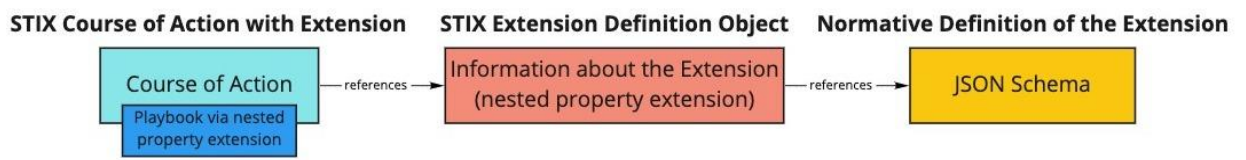

Figure 4. Course of Action instance that uses a nested property extension.

3. Define additional properties for an existing STIX Object type at the object's top-level. This can be done to represent properties that form an inherent part of the definition of an object type. Figure 5 presents a Course of Action instance with a top-level property extension.

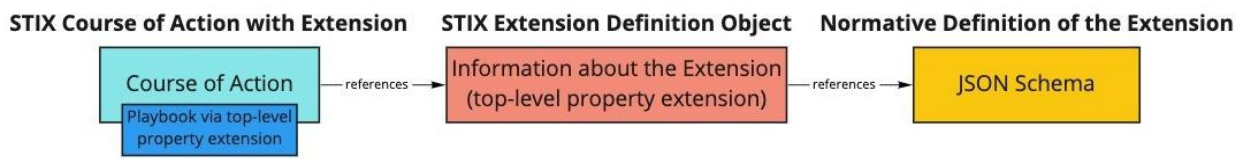

Figure 5. Course of Action instance that uses a top-level property extension.

## 9. Extending the Course of Action SDO Type to Share Security Playbooks

An SDO comprises a set of properties that, in combination, provide the essential context required to represent a concept. Relating the different concepts allows us to describe a threat landscape more comprehensively in a format that machines can understand and utilize. Similarly, extending a concept (using an *extension definition*) and, in our case, the Course of Action SDO to support sharing machine-readable security playbooks, requires introducing a set of meaningful properties that comprise the extension. Our requirements for the extension are to support sharing entire playbooks in their native or encoded format so that they can be easily shared and utilized programmatically and make available a set of metadata for descriptive and administrative purposes.

We found the work of Mavroeidis et al. [mezcj-2021] to satisfy our requirements. We utilized their proposed metadata template to introduce a nested property extension for the STIX Course of Action object to enable describing, embedding, and sharing security playbooks. The playbook metadata template of Mavroeidis et al. is presented in Table 4. The same table presents and describes the template's counterpart properties we propose and their data types that comprise the STIX *extension definition* for the Course of Action SDO. The *extension definition* defines a nested property extension that enables sharing security playbooks in STIX 2.1.

Table 4. Mavroeidis et al. [mezcj-2021] Security Playbook Metadata Template and STIX 2.1 Course of Action Extension Properties.

| Mavroeidis et al. | | STIX COA Property Extension | | |
|---|---|---|---|---|
| **Playbook Metadata** | **Description** | **Property Name** | **STIX Type** | **Description** |
| - | - | **extension_type** (required) | string | The value of this property **MUST** be `property-extension`. |
| **id** | A value that uniquely identifies the playbook. | **playbook_id** (optional) | string | A value that (uniquely) identifies the playbook. If the playbook itself embeds an identifier then the `playbook_id` **SHOULD** use the same identifier (value) for correlation purposes. |
| - | - | **created** (required) | timestamp | The time at which the playbook extension (instance) was created. This may be different than the time at which the "parent" Course of Action object instance was created. |
| - | - | **modified** (required) | timestamp | The time at which the playbook extension (instance) was last modified. A modification in the extension requires updating the property modified in both the extension and the "parent" Course of Action object. |
| **revoked** | A boolean that identifies if the playbook creator deems that this playbook is no longer valid. | **revoked** (optional) | boolean | A boolean that identifies if the playbook extension (instance) is no longer valid. |
| **creator** | The entity that created this playbook. It can be, for example, a natural person or an organization. It may be represented using an id that identifies the creator. | **playbook_ creator** (optional) | identifier | The identifier of the entity that created the playbook. |
| **created** | The time at which the playbook was originally created. | **playbook_ creation_time** (optional) | timestamp | The time at which the playbook was originally created. |
| **modified** | The time that this particular version of the playbook was last modified. | **playbook_ modification_ time** (optional) | timestamp | The time at which the playbook was last modified. |
| **valid_from** | The time from which the playbook is considered valid and the steps that it contains can be executed. | **playbook_ valid_from** (optional) | timestamp | The time from which the playbook is considered valid and the steps that it contains can be executed. |
| **valid_until** | The time at which this course of action/playbook should no longer be considered a valid playbook to be executed. | **playbook_ valid_until** (optional) | timestamp | The time from which the playbook should no longer be considered a valid playbook to be executed. |
| **description** | An explanation, details, and more context about what this playbook does and tries to accomplish. | **description** (optional) | string | An explanation, details, and more context about what this playbook does and tries to accomplish. |
| **label** | A set of terms, labels, or tags associated with this playbook (e.g., aliases of adversary groups or malware family/variant/name that this playbook is related to). | **labels** (optional) | list of type string | A set of labels for the playbook (e.g., adversary persona names, associated groups, or malware family/variant/name that this playbook is related to). |

| impact | From 0 to 100, an integer representing the impact the playbook has on the organization. A value of 0 means specifically undefined. Values range from 1, the lowest impact, to a value of 100, the highest. For example, a purely investigative playbook that is non-invasive would have a low impact value of 1. In contrast, a playbook that performs changes such as adding rules into a firewall would have a higher impact value. | `playbook_impact` (optional) | `integer` | From 0 to 100, an integer representing the impact the playbook has on the organization. A value of 0 means specifically undefined. Impact values range from 1, the lowest impact, to a value of 100, the highest. For example, a purely investigative playbook that is non-invasive could have a low impact value of 1. In contrast, a playbook that performs changes such as adding rules into a firewall should have a higher impact value. |
|---|---|---|---|---|
| severity | From 0 to 100, an integer representing the seriousness of the conditions that this playbook addresses. A value of 0 means specifically undefined. Values range from 1, the lowest severity, to a value of 100, the highest. | `playbook_severity` (optional) | `integer` | From 0 to 100, an integer representing the seriousness of the conditions that this playbook addresses. A value of 0 means specifically undefined. Severity values range from 1, the lowest severity, to a value of 100, the highest. |
| priority | From 0 to 100, an integer representing the priority of this playbook relative to other defined playbooks. A value of 0 means specifically undefined. Values range from 1, the highest priority, to a value of 100, the lowest. | `playbook_priority` (optional) | `integer` | From 0 to 100, an integer representing the priority of this playbook relative to other defined playbooks. A value of 0 means specifically undefined. Priority values range from 1, the highest priority, to a value of 100, the lowest. This property can support addressing different use cases and requirements of a producing or consuming entity. For |

| | | | | example, a STIX 2.1 COA object that embeds more than one playbook can use this property to define a priority with regards to which one should be executed. |
|---|---|---|---|---|
| **organization_ type** | The type of organization that the playbook is intended for. This can be an industry-sector. | **organization_ type** (optional) | `list` of type `open-vocab` | The type of organization that the playbook is intended for. The value for this property SHOULD come from the `industry-sector-ov` open vocabulary as defined in STIX Version 2.1. |
| **playbook_type** | The security-related functions the playbook addresses. A playbook may account for multiple types (e.g., detection and investigation). ['Notification', 'Detection', 'Investigation', 'Prevention', 'Mitigation', 'Remediation', 'Attack'] | **playbook_type** (optional) | `list` of type `open-vocab` | A list of playbook types that specifies the operational roles this playbook addresses. The open vocabulary is based on the available options in the CACAO standard and NIST SP 800-61 rev2.<br><br>Open Vocabulary: `[notification, detection, investigation, prevention, mitigation, remediation, analysis, containment, eradication, recovery, attack]` |
| **playbook_ standard** | The standard the playbook conforms to (e.g., CACAO). | **playbook_ standard** (optional) | `string` | The standard/format/notation the playbook conforms to (e.g., CACAO, BPMN). |
| **playbook_ abstraction** | The playbook's level of abstraction. ['Template', 'Executable'] | **playbook_ abstraction** (optional) | `open-vocab` | The playbook's level of abstraction (with regards to consumption). |

| | | | | Open Vocabulary: [template, executable] |
|---|---|---|---|---|
| **playbook** | The entire playbook in its native format (e.g., CACAO JSON). Security playbook producers and consumers use this property to share and retrieve playbooks. | - | - | - |
| **playbook_base64** | The entire playbook encoded in base64. Security playbook producers and consumers use this property to share and retrieve playbooks. | `playbook_bin` (optional) | `binary` | The entire playbook encoded in base64. Security playbook producers and consumers use this property to share and retrieve entire playbooks. |

An example STIX 2.1 Course of Action instance with a security playbook extension is presented below in JSON. Reexamining Figure 4, this instance represents the blue rectangle.

```json
{
    "type": "bundle",
    "id": "bundle--5e04bf76-5971-418e-b145-4dad3158e843",
    "objects": [
        {
            "type": "course-of-action",
            "spec_version": "2.1",
            "id": "course-of-action--e06259ad-a154-4e23-bc0a-e229ccb3456f",
            "created_by_ref": "identity--ae82a5e5-ec07-4863-ad88-6504b29f24e9",
            "created": "2022-01-18T23:22:03.934698Z",
            "modified": "2022-08-25T19:14:15.437976Z",
            "name": "playbook",
            "description": "A course of action for CVE-2021-44228.",
            "extensions": {
                "extension-definition--1e1c1bd7-c527-4215-8e18-e199e74da57c": {
                    "extension_type": "property-extension",
                    "playbook_id": "cf5997e8-e387-426a-a32d-694e4f55f80b",
                    "created": "2022-01-18T23:22:03.934698Z",
                    "modified": "2022-08-25T19:14:15.437976Z",
                    "playbook_creator": "identity--ae82a5e5-ec07-4863-ad88-6504b29f24e9",
                    "revoked": false,
                    "labels": [
                        "CVE-2021-44228"
                    ],
                    "description": "A playbook that, via SBOM processing, detects assets vulnerable to CVE-2021-44228. The same playbook will investigate if there have been attempts to exploit vulnerable assets.",
                    "playbook_valid_from": "2022-03-18T00:00:00.000000Z",
                    "playbook_creation_time": "2022-01-09T08:00:33.432637Z",
                    "playbook_impact": 1,
                    "playbook_severity": 90,
```

```
                "playbook_priority": 0,
                "playbook_type": [
                    "detection",
                    "investigation"
                ],
                "playbook_standard": "cacao",
                "playbook_abstraction": "executable",
                "playbook_base64": "U2VjdXJpdHkgUGxheWJvb2s="
            }
        }
    }
  ]
}
```

The Extension Definition object presented below (and referenced in the above example) contains information about the extension and provides a pivotal point (*schema* property) to retrieve the normative definition of the extension. Reexamining Figure 4, this object represents the red rectangle.

```
{
    "type": "bundle",
    "id": "bundle--68754162-c445-4996-bf3a-30b0ed54850d",
    "objects": [
        {
            "type": "extension-definition",
            "spec_version": "2.1",
            "id": "extension-definition--1e1c1bd7-c527-4215-8e18-e199e74da57c",
            "created_by_ref": "identity--ae82a5e5-ec07-4863-ad88-6504b29f24e9",
            "created": "2022-01-18T23:22:03.933931Z",
            "modified": "2022-08-25T19:15:25.577633Z",
            "name": "Course of Action extension for Security Playbooks",
            "description": "This extension definition extends the Course of Action SDO with additional properties for representing, managing, and sharing machine-readable security playbooks.",
            "schema": "https://raw.githubusercontent.com/fovea-research/stix2.1-coa-playbook-extension/main/schema/course-of-action_playbook.json",
            "version": "3.0.0",
            "extension_types": [
                "property-extension"
            ],
            "external_references": [
                {
                    "source_name": "GitHub",
                    "description": "Documentation of the Extension Definition.",
                    "url": "https://github.com/fovea-research/stix2.1-coa-playbook-extension"
                }
            ]
        }
    ]
}
```

The normative definition of the extension (JSON schema) is provided below. Reexamining Figure 4, this instance represents the amber rectangle.

```json
{
    "type": "bundle",
    "id": "bundle--c9661f66-3ff5-4431-9d52-64e18cc2c0b7",
    "objects": [
        {
            "type": "extension-definition",
            "spec_version": "2.1",
            "id": "extension-definition--1e1c1bd7-c527-4215-8e18-e199e74da57c",
            "created_by_ref": "identity--ae82a5e5-ec07-4863-ad88-6504b29f24e9",
            "created": "2022-01-18T23:22:03.933931Z",
            "modified": "2022-08-25T19:15:25.577633Z",
            "name": "Course of Action extension for Security Playbooks",
            "description": "This extension definition extends the Course of Action SDO with additional properties for representing, managing, and sharing machine-readable security playbooks.",
            "schema": "https://raw.githubusercontent.com/fovea-research/stix2.1-coa-playbook-extension/main/schema/course-of-action_playbook.json",
            "version": "3.0.0",
            "extension_types": [
                "property-extension"
            ],
            "external_references": [
                {
                    "source_name": "GitHub",
                    "description": "Documentation of the Extension Definition.",
                    "url": "https://github.com/fovea-research/stix2.1-coa-playbook-extension"
                }
            ]
        },
        {
            "type": "identity",
            "spec_version": "2.1",
            "id": "identity--ae82a5e5-ec07-4863-ad88-6504b29f24e9",
            "created": "2021-02-15T11:25:33.086853Z",
            "modified": "2021-07-02T10:57:28.592252Z",
            "name": "Vasileios Mavroeidis and Mateusz Zych",
            "identity_class": "group",
            "contact_information": "vas.mavroidis@gmail.com, zychmateusz93@gmail.com"
        },
        {
            "type": "course-of-action",
            "spec_version": "2.1",
            "id": "course-of-action--e06259ad-a154-4e23-bc0a-e229ccb3456f",
            "created_by_ref": "identity--ae82a5e5-ec07-4863-ad88-6504b29f24e9",
            "created": "2022-01-18T23:22:03.934698Z",
            "modified": "2022-08-25T19:14:15.437976Z",
            "name": "playbook",
            "description": "A course of action for CVE-2021-44228.",
            "extensions": {
                "extension-definition--1e1c1bd7-c527-4215-8e18-e199e74da57c": {
                    "extension_type": "property-extension",
                    "playbook_id": "cf5997e8-e387-426a-a32d-694e4f55f80b",
                    "created": "2022-01-18T23:22:03.934698Z",
                    "modified": "2022-08-25T19:14:15.437976Z",
                    "playbook_creator": "identity--ae82a5e5-ec07-4863-ad88-6504b29f24e9",
```

```
                    "revoked": false,
                    "labels": [
                        "CVE-2021-44228"
                    ],
                    "description": "A playbook that, via SBOM processing, detects assets vulnerable to CVE-2021-44228. The same playbook will investigate if there have been attempts to exploit vulnerable assets.",
                    "playbook_valid_from": "2022-03-18T00:00:00.000000Z",
                    "playbook_creation_time": "2022-01-09T08:00:33.432637Z",
                    "playbook_impact": 1,
                    "playbook_severity": 90,
                    "playbook_priority": 0,
                    "playbook_type": [
                        "detection",
                        "investigation"
                    ],
                    "playbook_standard": "cacao",
                    "playbook_abstraction": "executable",
                    "playbook_base64": "U2VjdXJpdHkgUGxheWJvb2s="
                }
            }
        }
    ]
}
```

The updated documentation, Extension Definition object, and schema that provides the normative definition of the extension are available on GitHub[3].

---

[3] https://github.com/fovea-research/stix2.1-coa-playbook-extension